\documentclass[pra,superscriptaddress,twocolumn,showpacs,floatfix]{revtex4}
\usepackage{bm}
\usepackage{graphicx} 

\begin{document}

\title{Symplectic tomography of ultracold gases in tight-waveguides}

\author{A. del Campo\footnote{E-mail: adolfo.delcampo@ehu.es}}
\affiliation{Departamento de Qu\'\i mica-F\'\i sica,
Universidad del Pa\'\i s Vasco, Apartado 644, Bilbao, Spain} 

\author{V. I. Man'ko\footnote{E-mail: manko@sci.lebedev.ru}}
\affiliation{P. N. Lebedev Physical Institute, Leninskii Prospect 53, Moscow 119991, Russia} 

\author{G. Marmo \footnote{E-mail: marmo@na.infn.it}}
\affiliation{Dipartamento di Scienze Fisiche, Universit\`{a} di Napoli ``Federico II'', I-80126 Naples, Italy} 
\affiliation{INFN, Sezione di Napoli, I-80126 Naples, Italy}

\def\M{{\rm M}}
\def\d{{\rm d}}
\def\t{{\rm t}}
\def\K{{\rm K}}
\def\la{\langle}
\def\ra{\rangle}
\def\om{\omega}
\def\Om{\Omega}
\def\vep{\varepsilon}
\def\wh{\widehat}
\def\tr{\rm{Tr}}
\def\da{\dagger}
\def\iz{\left}
\def\zi{\right}
\newcommand{\Xv}{\mbox{\boldmath$X$}}
\newcommand{\nuv}{\mbox{\boldmath$\nu$}}
\newcommand{\muv}{\mbox{\boldmath$\mu$}}
\newcommand{\beq}{\begin{equation}}
\newcommand{\eeq}{\end{equation}}
\newcommand{\beqa}{\begin{eqnarray}}
\newcommand{\eeqa}{\end{eqnarray}}
\newcommand{\intf}{\int_{-\infty}^\infty}
\newcommand{\into}{\int_0^\infty}

\begin{abstract} 

The phase space is the natural ground to smoothly extrapolate between local and non-local correlation functions.
With this objective, we introduce the symplectic tomography of many-body quantum gases in tight-waveguides, 
and concentrate on the reduced single-particle symplectic tomogram (RSPST) whose marginals are the density profile 
and momentum distribution. 
We present an operational approach to measure the RSPST from the time evolution of the density profile after shutting off the interactions
in a variety of relevant situations: free expansion, fall under gravity, and oscillations in a harmonic trap.
From the RSPST, the one-body density matrix of the trapped state can be reconstructed.

\end{abstract}

\maketitle

At low densities, ultracold bosonic gases exhibits universality. 
The interatomic interactions are then well described by the Fermi-Huang pseudo-potential, parametrized by a 
the 3D s-wave scattering length
$a_s$. If such gases are further confined in tight-waveguides, whenever the transverse excitation quantum 
$\hbar\omega_{\perp}$ is larger than the longitudinal zero point and 
thermal energies, the system effectively becomes one-dimensional \cite{Olshanii98}.
The interparticle pseudo-potential is then a simple  delta function, 
so the system is well approximated by the Lieb-Liniger model \cite{LieLin63}.
Moreover, the strength of the interaction as a function of $a_s$
 exhibits a confinement-induced 1D Feshbach resonance (CIR) \cite{Olshanii98,BerMooOls03}, 
allowing to tune the 1D coupling constant  $g_{1D}$ from $-\infty$
to $+\infty$ and to reach both weak and strongly interacting regimes \cite{PetShlWal00}. 
As a consequence, paradigmatic examples of the Bose-Fermi duality have been explored   
such as the Tonks-Girardeau gas \cite{exp}, in which the strongly repulsive interactions between 
bosons leads to an effective Pauli exclusion principle \cite{Girardeau60}.
In this regime,  the system undergoes fermionization, all local 
correlation functions being identical to those of the spin-polarized ideal Fermi gas.
Actually,  the Fermi-Bose duality comes into play even with finite interactions \cite{CS99}.
However, quantum statistics invariably imposes an underlying signature manifested when looking 
at non-local correlations such as the momentum distribution or the one-body density matrix 
\cite{RM05,MG05,Cazalilla02,BGOL05,chinos,ZMS06}.
A natural ground to smoothly extrapolate between local and non-local correlations is the phase space.
In this paper, we shall undertake the description of ultracold gases in tight-waveguides 
by means of the quantum tomographic technique \cite{MMT96}. 
We show that after shutting off the interactions in the system, 
the time evolution of the density profile governed by a quadratic Hamiltonian 
is tantamount to the knowledge of the (reduced) symplectic tomogram, 
from which the initial one-body density matrix of the trapped state 
can be reconstructed. This includes relevant experimental situations 
for ultracold atoms in waveguides 
such as  free expansion, dynamics falling under gravity, 
and time-evolution in a harmonic trap.

{\it Symplectic tomography.}  In the symplectic tomography probability representation, 
first introduced in \cite{MMT96}, the wave function $\Psi(z)$ 
or the density matrix $\rho(z,z')$ can be mapped onto the 
standard positive distribution ${\cal W}(X,\mu,\nu)$ 
of the random variable $X$ depending on two real
extra parameters, $\mu$ and $\nu$. The map is
given by the formula 
%
\begin{equation}\label{probdistribution}
    {\cal W}(X, \mu,\nu)= \frac{1}{2\pi\hbar|\nu|}\int\rho(z,z')
    e^{i\frac{\mu(z^{2}-{z'}^{2})}{2\hbar\nu}-i\frac{X}{\hbar\nu}(z-z')}\d z'
    \d z,
\end{equation}
which is the fractional Fourier transform \cite{Margarita00} of the density
matrix. The map is invertible so that the density matrix can be
expressed in terms of the tomographic probability representation
as follows,

\begin{equation}\label{densitytomograph}
    \rho(z,z')=\frac{1}{2\pi} \int
    \mathcal{W}(X,\mu,\frac{z-z'}{\hbar})\,
    e^{i\left(X-\mu\frac{z+z'}{2}\right)}\d X\d\mu.
\end{equation}
The expression in Eq. (\ref{probdistribution}) admits an affine
invariant form \cite{MMM01,MMM02}
\begin{equation}\label{invariantform}
  {\cal W}(X, \mu,\nu)={\rm Tr}[\hat{\rho} \delta(X-\mu\hat{z}-\nu\hat{p})],
\end{equation}
where  the density operator is denoted by
$\hat{\rho}$, and $\hat{z}$, $\hat{p}$
are the operators of position and the
conjugate momentum respectively. From Eq. (\ref{invariantform})
some properties of the tomogram ${\cal W}(X, \mu,\nu)$ are
easily extracted. First, the tomogram is a normalized
probability distribution, $\int{\cal W}(X,\mu,\nu) \d X=1$,
if the density operator is accordingly normalized (i.e. ${\rm Tr}\hat{\rho}=1$). 
Moreover, the tomogram satisfies
the homogeneity property \cite{MRV98}
\begin{equation}\label{homogeneity}
{\cal W}(\lambda X,\lambda\mu,\lambda\nu)= \frac{1}{|\lambda|}{\cal W}(X,\mu,\nu),
\end{equation}
inherited from that of the delta function in the definition, Eq. (\ref{invariantform}). 
This equation provides a link with optical tomography ($\mu=\cos\theta$, $\nu=\sin\theta$) 
\cite{BB87} and Fresnel tomography ($\mu=1$) \cite{DFMM05}. 
We further notice that the tomogram can be related to the
Wigner function $W(z,p)$ \cite{Wigner32},
\begin{equation}\label{wignertomo}
  {\cal W}(X,\mu,\nu)= \int W(z,p)\delta(X-\mu z - \nu p)\frac{\d z \d p}{2\pi\hbar}
\end{equation}
as its Radon transform, providing a clear interpretation of the tomogram $ {\cal W}(X,\mu,\nu)$. 
One has the line $X=\mu z + \nu p$ in phase space, which is given by equating to zero 
the delta-function argument in
Eq. (\ref{wignertomo}). Alternatively, the parameters $\mu$ and $\nu$
can be expressed in the form $s\cos\theta$, $\nu=s^{-1}\sin\theta$.
Here $s>0$ is a real squeezing parameter and $\theta$ is a rotation
angle. Then the variable $X$ is identical to the position measured
in the new reference frame in the phase-space. The new
reference frame has new scaled axis $sz$ and $s^{-1}p$ and after
the scaling the axis are rotated by an angle $\theta$. Thus the
tomogram implies the probability distribution of the random
position $X$ measured in the new (scaled and rotated) reference
frame in the phase-space. The remarkable property of the
tomographic probability distribution is that it is  a fair positive
probability distribution and it contains a complete information of
the state. Indeed, the  density operator
$\hat{\rho}$ can be expressed in terms of the tomogram as
\cite{DMMT98}
\begin{equation}\label{operatorrho}
\hat{\rho}=\frac{\hbar}{2\pi}\int\mathcal{W}(X,\mu,\nu)e^{i(X-\mu\hat{z}-\nu\hat{p})}
\d X \d\mu \d\nu.
\end{equation}
The tomographic map can be used not only for the
description of the state in terms of a probability distributions,
but also to describe its evolution (quantum
transitions) by means of the standard real positive transition
probabilities (alternative to the complex transition probability amplitude), 
i.e., in tunnelling dynamics \cite{LSA04}.

{\it Many-body 1D gases.} We next focus on effectively one-dimensional many-body systems, 
described generally by a wavefunction $\psi(z_1,\dots,z_N)$ 
or alternatively the N-body density matrix $\rho_N(z_1,\dots,z_N,z_1',\dots,z_N')$.
Introducing the notation 
${\bf z}=z_1,\dots,z_N$, 
and similarly for 
$\{\Xv, \nuv, \muv\}$,
one finds the many-body tomogram carrying out the $2N$-dimensional integral,

\beqa
& & {\cal W}_N(\Xv, \muv,\nuv)= \frac{1}{(2\pi\hbar)^N \prod_i|\nu_i|}
\int\rho_N({\bf z},{\bf z}')\nonumber\\
& & \times e^{i\sum_{j}\big[\frac{\mu_j(z_{j}^{2}-z_{j}'^{2})}{2\hbar\nu_j}-\frac{X_j}{\hbar\nu_i}(z_j-z'_j)\big]}
\d {\bf z}\d {\bf z}'.
\eeqa
Let us define the reduced single-particle symplectic tomogram (RSPST) as
\beqa
{\mathcal W}(X,\mu,\nu)=\int\prod_{j=2}^N\d X_j {\cal W}_N(\Xv, \muv,\nuv).
\eeqa
From the basis invariance property of the trace, it follows that alternatively one can find 
${\mathcal W}(X,\mu,\nu)$ through Eq. (\ref{probdistribution}), using the reduced single-particle density matrix (RSPDM) 
$\rho(z,z')\!=\!\int\!\d z_2\dots\d z_N 
\psi(z,z_2,\dots,z_N)\psi(z',z_2,\dots,z_N)^*$. Note that we choose the normalization $\int\d z\rho(z,z)=1$. 
Further notice the normalization condition $\int {\mathcal W}(X,\mu,\nu) \d X=1$, 
and that the symplectic tomogram satisfies the following two marginals of the Wigner function, 
 \beqa
 \label{rhox}
 {\mathcal W}(z,1,0)=n(z),~~~~~~~~~{\mathcal W}(p,0,1)=\varrho(p),
 \eeqa
where $n(z)=\rho(z,z)$ is the density profile, 
and $\varrho(p)=(2\pi\hbar)^{-1}\int\d z\d z'\rho(z,z')e^{ip(z'-z)/\hbar}$ 
 the momentum distribution.
Therefore, the  reduced tomogram extrapolates smoothly between the (local) density 
profile and the (non-local) momentum distribution performing a rotation in phase space 
parametrized by $(\mu,\nu)$.

{\it Measuring the RSPST.} A direct experimental measurement of the matter-wave symplectic tomograms has been eluded so far. 
To close this gap, we shall consider different situations to implement physically the symplectic tomography representation. 
We shall next focus on the Lieb-Liniger model \cite{LieLin63} which accurately describes ultracold 
atom vapors strongly confined in waveguides \cite{Olshanii98,BerMooOls03,PetShlWal00}.
The Hamiltonian is that of $N$ trapped bosons with pairwise delta interactions, 
\beqa
\mathcal{H}_{LL}=\sum_{i=1}^N H_i +g_{1D}(t)\sum_{1\leq i<j\leq N}
\delta(z_i-z_j), 
\eeqa
 where $H_i=-\frac{\hbar^2}{2m}\frac{\partial^2}{\partial z_{i}^2 }+V(z_i)$ is the single-particle Hamiltonian, 
and $V(z)$ denotes the trapping potential.
The value of $g_{1D}$ is a function of $a_s$ with a confinement-induced 1D Feshbach resonance \cite{Olshanii98,BerMooOls03}, 
which allows to tune  $g_{1D}$ from $-\infty$ to $+\infty$ by shifting the position of a 3D Feshbach scattering resonance
 via an external magnetic field \cite{Rob01}.

In what follows we shall consider the time-dependence of the coupling strength to be $g_{1D}(t)=g_{1D}\Theta(-t)$  
(where $\Theta(t)$ is the Heaviside step function), so that the interactions are turned off for $t>0$.
The RSPST can then be related to the dynamics governed by an external potential.

If for $t>0$ the trap is also turned off the Hamiltonian becomes the kinetic energy operator, 
and the system undergoes free expansion.
The free time evolution is governed by the kernel \cite{GS98}
\beqa
\label{rdmprop}
\mathcal{K}_0(z,z',t|z_0,z_0',0)=\la z_0'|\hat{U}_0^{\dag}(t)|z'\ra\la z|\hat{U}_0(t)|z_0\ra
\eeqa
 where the matrix elements (propagator) of the free evolution operator $\hat{U}_0(t)$ read
\beqa
\label{freeprop}
\la z|\hat{U}_0(t)|z_0\ra
=\sqrt{\frac{m}{2\pi i\hbar t}}e^{i\frac{m(z-z_0)^2}{2\hbar t}},
\eeqa
so that 
\beqa
\rho_{0}(z,z',t)=\!\int\!\d z_0\d z_0'\mathcal{K}_0(z,z',t|z_0,z_0',0)\rho(z_0,z_0').
\eeqa	
It follows that the RSPST can be obtained from 
 the free propagation of the density profile $n_0(z,t)=\rho_{0}(z,z,t)$,
\beqa
\label{free}
{\cal W}(X, \mu,\nu)=\frac{1}{|\mu|}n_{0}\left(\frac{X}{\mu},\frac{m \nu}{\mu}\right).
\eeqa	
The expansion takes place here in one dimension rather than three, this is, 
while keeping the radial confinement on. We further notice that if interactions are kept, 
important deviations from the ballistic expansion may occur \cite{OS02,DM06}, 
and the Lieb-Liniger dynamics blurs the information of the initial state \cite{Girardeau03,BPG07}.

A careful analysis  in the presence of the gravitational potential $V(z)=mgz$, 
whose propagator can be related to the free one in Eq. (\ref{freeprop}) as $\la z|\hat{U}_0(t)|z_0\ra{\rm exp}(-img(z+z_0)t/2\hbar-im^2g^2t^3/24\hbar)$  \cite{GS98}, allows to obtain the RSPST 
in an analogously way from the density profile, with the replacement $
z\rightarrow X/\mu-gm^2\nu^2/2\mu^2$.
Therefore, the expansion both in a horizontal or vertical waveguide allows to reconstruct 
the reduced tomogram and hence the one-body density matrix of the initial state.

Alternatively, the RSPST can be measured in a harmonic trap without releasing the gas, 
just by shutting off the interactions.
In this case the tomographic kernel in Eq. (\ref{probdistribution}) is implemented 
using the propagator \cite{GS98}
\beqa
\la z|\hat{U}_{trap}(t)|z_0\ra
=\sqrt{\frac{m\om}{2\pi i\hbar\sin\om t}}e^{i\frac{m\om \cot\om t}{2\hbar}(z^2+z_{0}^{2})
-i\frac{m \om z z_0}{\hbar\sin\om t}},\nonumber\\
\eeqa
whence it follows that
\beqa
{\cal W}(X,\mu,\nu)=\frac{1}{\lambda}n_{trap}\left(\frac{X}{\lambda},\frac{1}{\om}\tan^{-1}\frac{ m\om\nu}{ \mu}\right).
\eeqa	
with $\lambda=\sqrt{\mu^2+m^2\om^2\nu^2}$.
At variance with the operation approach based on free expansion, the dynamics is periodic, 
and it is possible to reconstruct the RSPST from the density profile along one cycle, $T=2\pi/\om$.

{\it Reconstruction of the density matrix.} The reconstruction of a quantum state from the dynamics of wavepackets in a potentail $V(z)$ 
has been successfully addressed by Leonhardt and coworkers within the  optical tomography \cite{LR96, LS97}.
Moreover, the Wigner function has been experimentally measured 
by looking at the time-evolution of the density profile of Helium atoms \cite{KPM97}; and recently, 
a scheme for neutron wavepacket tomography has been proposed \cite{neutron}.

In the following, we provide the explicit expressions for the density matrix corresponding to the situations 
discussed in the previous section.
Note that using Eqs. (\ref{invariantform}) and (\ref{free}), 
the free time evolution of the density profile can be written in the invariant form, 
 $n_0(z,t)=\la z|\hat{U}_0(t)\hat{\rho}\hat{U}_{0}^{\dag}(t)| z\ra
={\rm Tr}\big[\hat{\rho}\delta\!\big(z-\hat{z}-t\hat{p}/m\big)\big]$, 
which suggests that this type of reconstruction 
can be applied to other observables different from $\hat{\rho}$. 
Moreover, from Eq. (\ref{operatorrho}) and (\ref{free}), defining $X=\kappa z$, using the Baker-Campbell-Hausdorff formula and the Jacobian $|\partial(X,\nu,\mu)/\partial(\kappa,z,t)|$, the inverse reads
\beqa
\hat{\rho}=\frac{\hbar}{2\pi m}\int n_0(z,t)|\kappa|e^{i\kappa\left(z-\hat{z}-\frac{t}{m}\hat{p}\right)}\d \kappa\d z\d t.
\eeqa
Let as denote the Fourier transform of the density profile by 
$\tilde{n}_0(k,t)=\frac{1}{\sqrt{2\pi}}\int n_0(z,t)e^{-ikz}\d z$. 
In the wavevector $k$-representation one finds that the density matrix of the initial state is given by
\beqa
\label{rhokk}
\rho(k,k')\!=\!\frac{\hbar}{\sqrt{2\pi}m}\!\int\! \tilde{n}_0(k-k',t)|k-k'|e^{i\frac{\hbar(k^2-k'^2)t}{2m}} \d t, 
\eeqa
which is analogous to the equation derived in \cite{LS97} for wavepacket state reconstruction.
The Riemann-Lebesgue lemma limits in practice the range of the integral.
Alternatively, we note that the RSPDM can be diagonalized \cite{DG03} as $\rho(z,z';t=0)=
\sum_j\lambda_j\varphi_j(z)\varphi_j^{*}(z')$ in terms of the 
orthonormal natural orbitals $\varphi_j(z)$ with occupation numbers $\lambda_j>0$ 
satisfying $\int\rho(z,z')\varphi_j(z')dz'=\lambda_j\varphi_j(z)$ and $\sum_{j=1}^N\lambda_j=1$. 
Under free evolution, having set up $g_{1D}(t>0)=0$, the density profile reads 
$n_0(z,t)=\sum_j\lambda_j|\varphi_j(z,t)|^2$ with $\varphi_j(z,t)=(2\pi)^{-1/2}\int dk
\tilde{\varphi}_j(k)e^{ikz-i\hbar k^2 t/2m}$, which using Eq. (\ref{rhokk}), 
leads to the density matrix $\rho(k,k')=\sum_j\lambda_j\tilde{\varphi}_j(k)\tilde{\varphi}_j^{*}(k')$ \cite{comment}. 

When the density profile corresponds to the expansion falling under gravity, 
the above expression Eq. (\ref{rhokk}) holds  
with the definition $\tilde{n}_0(k,t)=\frac{1}{2\pi}\int n_0(z-gt^2/2,t)e^{-ikz}\d z$.

The dynamics in a harmonic trap after shutting off the interactions similarly allows 
to reconstruct the density matrix operator according to
\beqa
\hat{\rho}=\frac{\hbar}{2\pi m}\int n_{trap}(z,t)|\kappa|e^{i\kappa\left(z-\cos\om t\hat{z}-\frac{\sin\om t}{m\om}\hat{p}\right)}\d \kappa\d z\d t,\nonumber\\
\eeqa
or in the wavenumber representation,
\beqa
\rho(k,k')&=& \frac{\hbar}{\sqrt{2\pi}m}\int\frac{1}{\cos^2\om t} \tilde{n}_{trap}\left(\frac{k-k'}{\cos\om t},t\right)\nonumber\\
& & \times |k-k'|e^{i\frac{\hbar\tan\om t}{2m\om}(k^2-k'^2)}\d t,
\eeqa
where the caustic at half period is a well-known property of the Radon transform.
We note also that the obtained results can be put in the compact form 
\beqa
\hat{\rho}=\frac{\hbar}{2\pi m}\int n_{H}(z,t)|\kappa|e^{i\kappa(z-\hat{z}(t))}\d \kappa\d z\d t,
\eeqa
where $\hat{z}(t)=\hat{U}_{H}^{\dag}\hat{z}\hat{U}_H$, and $n_{H}(z,t)$ is the density profile evolving under the action of $\hat{U}_H$, 
the evolution operator corresponding to a quadratic Hamiltonian $H$ of free motion, gravitational fall, and harmonic oscillation.
This equation suggests that the reconstruction of the density matrix can be generalized 
for time dependent single particle quadratic Hamiltonians.

{\it Conclusions.} In conclusion, we have introduced the reduced symplectic tomography of many-body quantum gases confined in tight-waveguides.
By performing a rotation in phase space, a smooth extrapolation is possible between the (local) density profile  
and the (non-local) momentum distribution. 
Moreover, a simple procedure has been given to experimentally measure the 
 tomogram, namely, by suddenly shutting off the interactions and registering the time evolution 
of the density profile 
in a variety of situations: free expansion, expansion under gravity, 
and periodic motion in a harmonic trap. We note that such dynamics can only be implemented 
if the interparticle interactions are negligible. 
Else, the so-called cusp condition imposed on the wavefunction by the short-range pseudo-potential affects the density profile \cite{OS02,BPG07} 
blurring the information of the initial state. 
In addition, it should be clear that even though the density profile of duals systems within the trap is exactly the same, 
once the interactions are switched off, the time evolution will differ provided that the momentum distribution is unlike in each system and therefore our method account for a proper description of different quantum statistics.
The one-body density matrix of the  state in the trap can then be reconstructed.
As long as the dynamics is governed by a quadratic Hamiltonian, the reconstruction is 
expected to be possible even for time-dependent potentials.
We close by noting that higher order correlations of the trapped gas can be inferred 
from the time evolution of the density profile \cite{ADM04}.

\begin{acknowledgments}

A. C. and V. I .M. thank the Universit\`{a} Federico II di Napoli for kind hospitality
during the completion of this work. A. C. further acknowledges discussions with J. G. Muga, 
and a fellowship from the Basque Government (BFI04.479) and the ESF (MISGAM-2116). 
This work has been supported by Ministerio de Educaci\'on y Ciencia (BFM2003-01003), 
and the Russian Foundation for Basic Research (07-02-00598).
\end{acknowledgments}


\begin{thebibliography}{10}
\bibitem{comment} As it is well known, assumption of time-reversal invariance or use of spatial symmetries of the problem at 
hand are required to extend the integral in time over the whole real line \cite{KPM97}.
\end{thebibliography}
\end{document}